\documentclass[preprint,authoryear,12pt]{elsarticle}

\usepackage{amssymb}

\usepackage[authoryear]{natbib}

\usepackage[
breaklinks=true,%
colorlinks=true,%
pdfauthor={Fainshtein V.G., Egorov Ya.I., Zagainova Yu.S.},%
pdftitle={KINEMATICS OF CMEs AND ASSOCIATED SHOCK WAVES AS DEDUCED FROM LASCO DATA: COMPARATIVE ANALYSIS}%
]{hyperref}

\journal{Advances in Space Research}

\begin{document}

\begin{frontmatter}



\title{Kinematics of CMEs and associated shock waves as deduced from LASCO data: comparative analysis}


\author{Fainshtein V.G.\corref{cor}}
\address{ISTP SB RAS, p.o.b. 291, Irkutsk, Russia}
\cortext[cor]{Corresponding author}
\ead{vfain@iszf.irk.ru }


\author{Egorov Ya.I.}
\address{ISTP SB RAS, p.o.b. 291, Irkutsk, Russia}
\ead{egorov@iszf.irk.ru }

\author{Zagainova  Yu.S.}
\address{IZMIRAN, Moscow, Troitsk}

\begin{abstract}

From data by LASCO C2 and C3 coronagraphs, depending on time (distance), we have determined positions and velocities of the front for fast limb CMEs' body with their sources near the limb, and for the body of halo-type CME with the sources near the solar disk center. These characteristics of CME body are compared to similar kinematic characteristics obtained for CME body-associated shock waves (shocks). For the body of halo-type CME with the sources near the solar disk center and associated shocks, we determined and compared their kinematic characteristics in 3D space. It has been shown that for all the considered CME groups, the shock velocity is higher than the CME body velocity, both velocities decrease as the mass ejection moves. As this takes place, the distance between CME body and shock grows. On average, distance from CME body to shock, and velocity difference of these structures is greater for a halo CME, and even greater for a model CME in 3D.

\end{abstract}


\end{frontmatter}

\parindent=0.5 cm

\section{Introduction}

It is now considered well-ascertained that in front of the fast coronal mass ejections (CME) observed in the field of view of LASCO C2 and/or C3 and other coronagraphs, there exist shocks \citep{Vourlidas2003,Ontiveros2009,Vourlidas2009,Vourlidas2012}. Today, such shocks are referred to as CME-driven shocks. In many cases, shocks in front of CMEs in coronagraph's field of view (FOV) can be visually seen as a distinct but faint brightness front of the external edge of the low-emission region in front of the CME bright frontal structure. The CME part, which is limited by the outer border of the frontal structure, we will call a CME body, though it is often referred to as a CME. This designation we will also use. The corona region between the frontal structure outer boundary and the shock is called “sheath”. Conceptually, it represents shock-compressed plasma behind the shock front.
Already visual observation of the CME body and associated shock movement in the corona white light images demonstrates that kinematics of these structures differs markedly. The shock is moving faster compared to the CME body, and the distance between them grows with time. In several works, a quantitative analysis of the two structures' kinematics has been carried out for several events based on observation analysis. Using direct observations of CMEs and CME-driven shock in FOV of COR2 coronagraphs and Heliospheric Imager 1 instruments of the Sun Earth Connection Coronal and Heliospheric Investigation suite \citep{stereo} on Solar Terrestrial Relations Observatory \citep{2008SSRv..136....5K}, \citet{2011ApJ...736L...5M} compared the time dependences of the CME front and associated shock positions for the 5 April 2008 event. It was shown that the altitude difference between the two structures increases with time up to ≈ 1/2AU. In \citep{Fainshtein2015}, using the CME “Ice-cream cone model” and based on CME 3D parameters calculation technique, proposed in \citep{Xue2005}, for several CMEs observed in FOV of the Large Angle and Spectrometric Coronagraph (LASCO; \citet{1995SoPh..162..357B}) onboard Solar and Heliospheric Observatory (SOHO; \citet{1995SoPh..162....1D}), the authors compared positions and velocities of the model CME body boundary and associated shock along the direction CME moves. It is shown that CME body and shock velocity decreases with time (with distance), and this process is faster for the body of mass ejection as compared to shock waves. Therewith, the velocity difference between the structures grows with time. The distance between model CME front and shock also increases with time. For the 03 April 2010 event, \citet{2016AIPC.1714c0003V} found the position and velocity dependencies separately for CME and associated shock using the STEREO data. The authors have found that with time, the distance between CME front and shock increases, specifically, in FOV of STEREO/COR2 coronagraphs, and velocities of both structures in COR2 FOV go down.
Note that in coronagraphs' FOV, many works estimated the $\Delta R$ parameter - shock standoff distance (distance between a CME body and associated shock on a CME axis), or standoff distance ratio $\Delta R/R_c$ (the standoff distance normalized by the radius of curvature of a CME (Rc)) (for instance, see \citet{2011ApJ...736L...5M,Gopalswamy2011,2012ApJ...759..103S,Fainshtein2015,2016AIPC.1714c0003V} and references therein). In \citep{Gopalswamy2011}, this parameter was derived from measurements in the FOV of LASCO coronagraphs for limb CMEs, and in \citep{2012ApJ...759..103S} - from 2.5D calculations of model CME movement in the form of flux rope and associated shock. The main reason of special interest in the standoff distance ratio is that the parameter is associated with the density compression ratio at the shock front and with Mach number for shock waves. It has been shown in \citep{2011ApJ...736L...5M,Fainshtein2015} that in COR2 and LASCO FOV, the shock standoff distance decreases as the CME moves. Though at great distances in the interplanetary space, this parameter grows with distance \citep{2011ApJ...736L...5M,2012ApJ...759..103S}. 
Movement of the CME body consists of two components. First, it moves translatory as a whole. Second, it expands maintaining roughly its angular sizes. Motion of the body front boundary of CMEs, whose sources are located near the solar limb, is the sum of translatory motion and expansion of the mass ejection. Movement of the CME body boundary, whose sources are located near the solar disk center, basically reflects extension of the CME body, and the shock registered in this case represents the flank areas of the whole CME-associated shock. These facts were virtually unaccounted in most of previous studies, which compared the CME and shock kinematics. The purpose of this paper is to compare kinematic characteristics (position and velocity) of fast CME bodies and associated shock waves based on data from the SOHO/LASCO-C2, C3 coronagraphs for two types of coronal mass ejection: limb CMEs and CMEs with their sources near the solar disk center; also, using the "Ice cream cone model" CME \citep{Xue2005} kinematic characteristics of a body of halo CME and its associated wave in 3D space must be compared with the results for limb CMEs and halo CMEs that have sources near the solar disk center, according to observations in the plane of the sky.

\section{Data and Methods of their Analysis}

We used LASCO C2, C3 coronagraphs data with Level-1 of image processing (\url{https://sharpp.nrl.navy.mil/cgi-bin/swdbi/lasco/images/form}). For analysis, from catalogs \url{https://cdaw.gsfc.nasa.gov/CME_list/} and \url{https://cdaw.gsfc.nasa.gov/CME_list/HALO/halo.html} we selected 11 limb CMEs with their sources within 30$^{\circ}$ related to the limb and 11 CMEs \& all halo CMEs) with their sources within 30$^{\circ}$ as related to the solar disk center.

\begin{table}[!ht]
\caption{left column: CMEs with their sources within 30$^{\circ}$ related to the limb; right column: halo CMEs) with their sources within 30$^{\circ}$ as related to the solar disk center} 
\centering 
\begin{tabular}{rrr|rrr} 
\hline\hline 
Date & Time & Velocity, km/s & Date & Time & Velocity, km/s\\[0.5ex]

\hline  
1998.04.20 & 10:07 & 1863 & 2001.04.10 & 05:30 & 2411\\[1ex] 1998.11.24 & 02:30 & 1799 & 2001.09.24 & 10:30 & 2402\\[1ex] 1999.06.11 & 01:26 & 1719 & 2003.10.28 & 11:30 & 2459\\[1ex] 1999.07.25 & 13:12 & 1389 & 2003.11.18 & 08:50 & 1660\\[1ex] 2001.12.14 & 09:06 & 1506 & 2004.11.07 & 16:.54 & 1759\\[1ex] 2004.07.29 & 12:30 & 1468 & 2005.01.15 & 06:30 & 2049\\[1ex] 2005.06.03 & 12:32 & 1679 & 2005.01.16 & 00:18 & 2861\\[1ex] 2005.07.27 & 05:08 & 1787 & 2005.09.13 & 20:00 & 1866\\[1ex] 2005.08.22 & 17:30 & 2378 & 2012.01.23 & 03:48 & 2175\\[1ex] 2005.08.23 & 14:40 & 1929 & 2012.03.07 & 00:36 & 2684\\[1ex] 2005.09.05 & 09:48 & 2326 & 2014.01.07 & 05:30 & 1830\\[1ex]
\hline  
\end{tabular} 

\label{tab1} 
\end{table} 
 The dates of events are shown, parentheses next to the event date include the time when mass ejection was first observed in LASCO C2 FOV. Next to these parentheses, we indicated the CME linear projection velocity in km/s, which exceeded 1500 km/s for all the selected events. The main criteria for the event selection is the presence of a clearly observed area of the shock-compressed plasma behind the shock front in the form of a diffuse faint-luminous area (see Fig.1) and, in most events, the possibility to determine a shock as a brightness jump (see below). 
The limb CMEs tend to move roughly perpendicular to the line of sight. In this case, according to observations in the coronagraph's FOV, the CME body is involved in two types of motion: translatory motion of the CME body as a whole and simultanous expansion of it, and specifically along the direction of its translatory motion. Herewith, due to their associated shock waves, some limb CMEs turn out to be halo CMEs at the same time, when a shock covers the coronagraph occulting disk all around (most often, such limb CMEs are observed in such a way only in LASCO C3 FOV). For halo CMEs with sources near the center of solar disk, movement of the CME body within the coronagraph's FOV primarily displays its transverse extension, and a shock wave is represented with its own flank areas.   
Fig. 1 demonstrates examples of the analyzed CMEs and shows the boundary of CME body and shock. 

\begin{figure}[!ht]
    {\includegraphics[width=1.\textwidth]{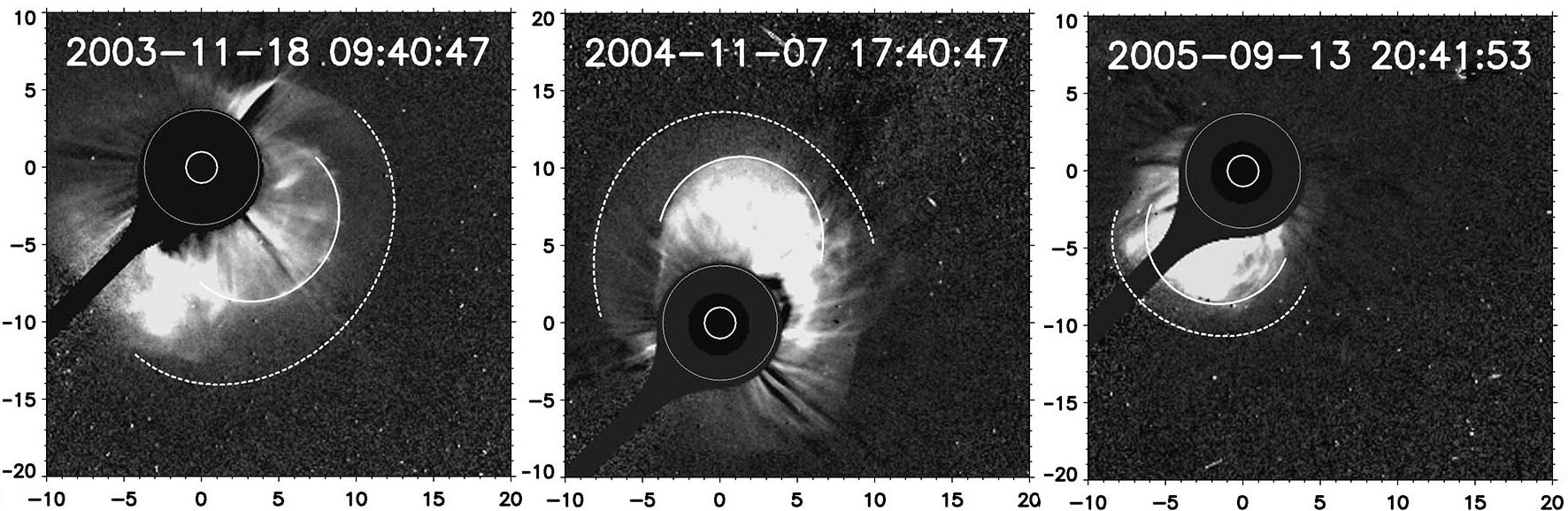}}
    \caption{Examples of limb and halo CMEs. White solid and dotted curves indicate the boundary of CME bodies and shock waves respectively. }
\end{figure}

To determine position of the CME boundary and shock in LASCO C2, C3 FOV, we used two methods: visual based on the corona images as a distinct boundary of the brightest structure (CME body) and less bright diffuse area (see Fig. 1), and by scanning the brightness of corona differential images (Fig. 2). In the first case, exact position of the visually selected points was calculated using special software that determined distance from the selected point to the solar disk center in solar radii. In the second case, we performed averaging over a different number of pixels along the white corona brightness scanning line, depending on coronagraph (LASCO C2 or C3), and on the noise level in corona image. Despite averaging, the width of the shock front did not exceed two spatial resolutions of coronagraph ($2\Delta$), where $\Delta = 0.025R_{\odot}$ for LASCO C2 and $\Delta = 0.125R_{\odot}$ for LASCO C3.  Scanning direction of white corona brightness was selected in such a way that it crossed the boundary of CME body and the clearly observed diffuse area behind the shock front at the same time. 

\begin{figure}[!ht]
    {\includegraphics[width=1.\textwidth]{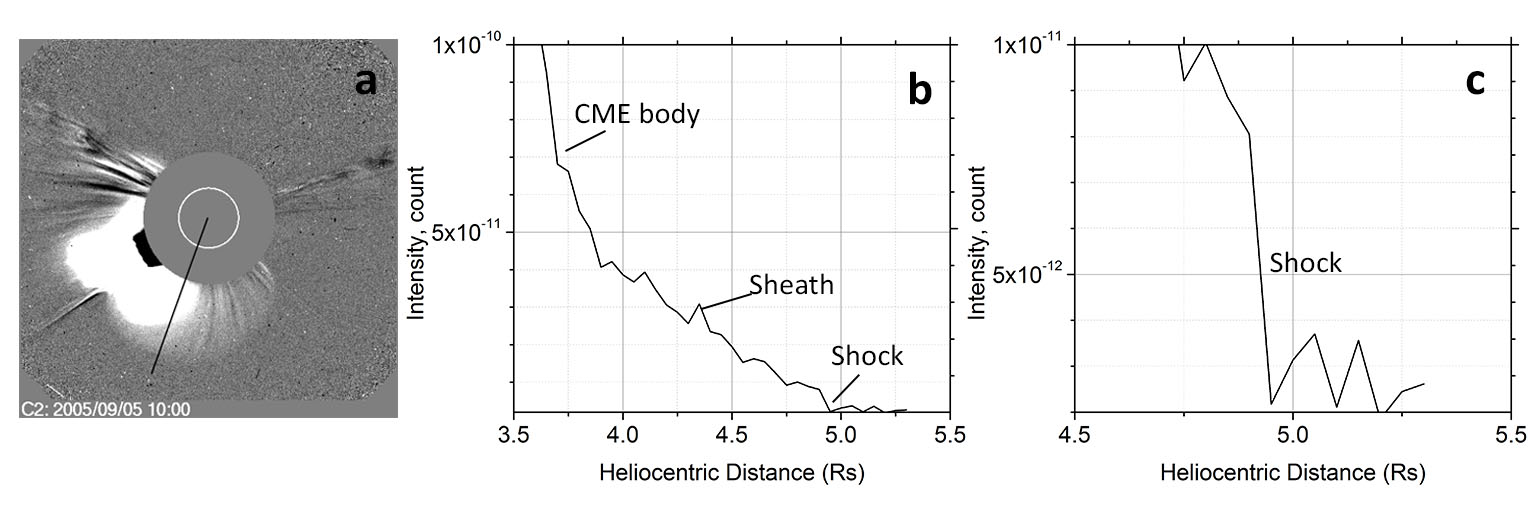}}
    \caption{(a) - LASCO C2 running difference image; (b) - brightness distribution (along black line shown on (a)), the boundary of CME body and shock shown. (c) - scaled plot b}
\end{figure}

In addition to comparing the kinematics of the CME body and associated shock in LASCO C2 and C3 FOV for limb CMEs and halo CMEs with the source center near the solar disk center, we compared kinematics of these structures to kinematics of CME body and shock in three dimensions using the "Ice-cream cone model" for a CME and the method of calculating a model 3D parameters proposed in \citep{Xue2005}.  The CME "Ice-cream cone model" is a cone with its apex in the center of the Sun, footpoint of which rests on a part of the sphere. The method proposed in \citep{Xue2005} allows determining of the following 3D parameters of a CME: direction of CME movement, CME angular size and position or velocity of the CME top at a fixed instant of time.
Peculiarity of using this method in our work is that we applied it separately for shock and boundary of CME body. Figure 3 shows the examples of selecting boundary fragments of the CME body and the shock that were used to calculate 3D parameters of the CME body and associated shock wave. In the left panel, fragments of the CME body boundary and shock are marked as separate points. This is how \citep{Xue2005} showed the CME outer boundary (normally it was a shock). We used yet one method to detect the boundary of CME body and shock, by placing ellipses with selected parameters on them. Although the results of calculating the CME and shock parameters in 3D space were close in both cases, the second way of detecting the structures in question allowed us to noticeably accelerate calculations of CME body and shock kinematic characteristics in 3D space. 
  
\begin{figure}[!ht]
    {\includegraphics[width=1.\textwidth]{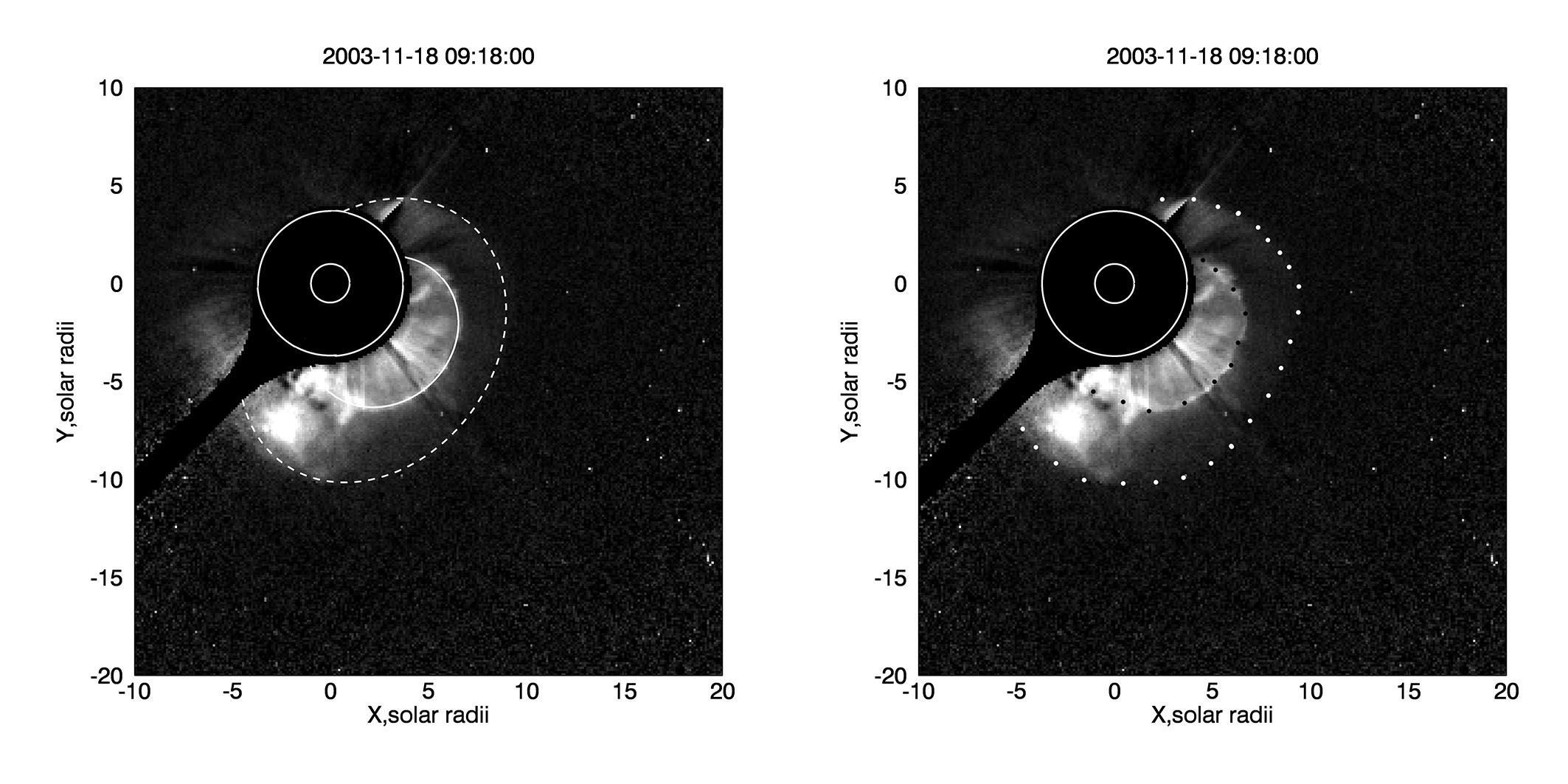}}
    \caption{Bright area is a CME body; diffusion area of low brightness that surrounds the CME body is the shock-compressed plasma behind the shock front. Points or white lines at the boundaries of these areas map out the sections that were used to calculate 3D parameters of CME body and shock.}
\end{figure}

Unlike our work \citep{Fainshtein2015}, here, to calculate 3D parameters of a CME body and a shock, we have selected only 5 halo CMEs with over 1500 km/s velocity, for which we calculated CME body and shock 3D parameters in a most reliable way.

\section{Results}
Figure 4 demonstrates kinematic characteristics (position and velocity) separately for the front of CME body and shock along one of directions that cross the external boundary of CME body and shock for the limb CME observed on 27.07.2005. We can see that over time, distance between shock and CME body grows, while at the same time velocity of CME body and shock goes down, but the shock velocity decreases faster. As a result, difference in velocities of shock and CME body decreases with time.

\begin{figure}[!ht]
    {\includegraphics[width=1.\textwidth]{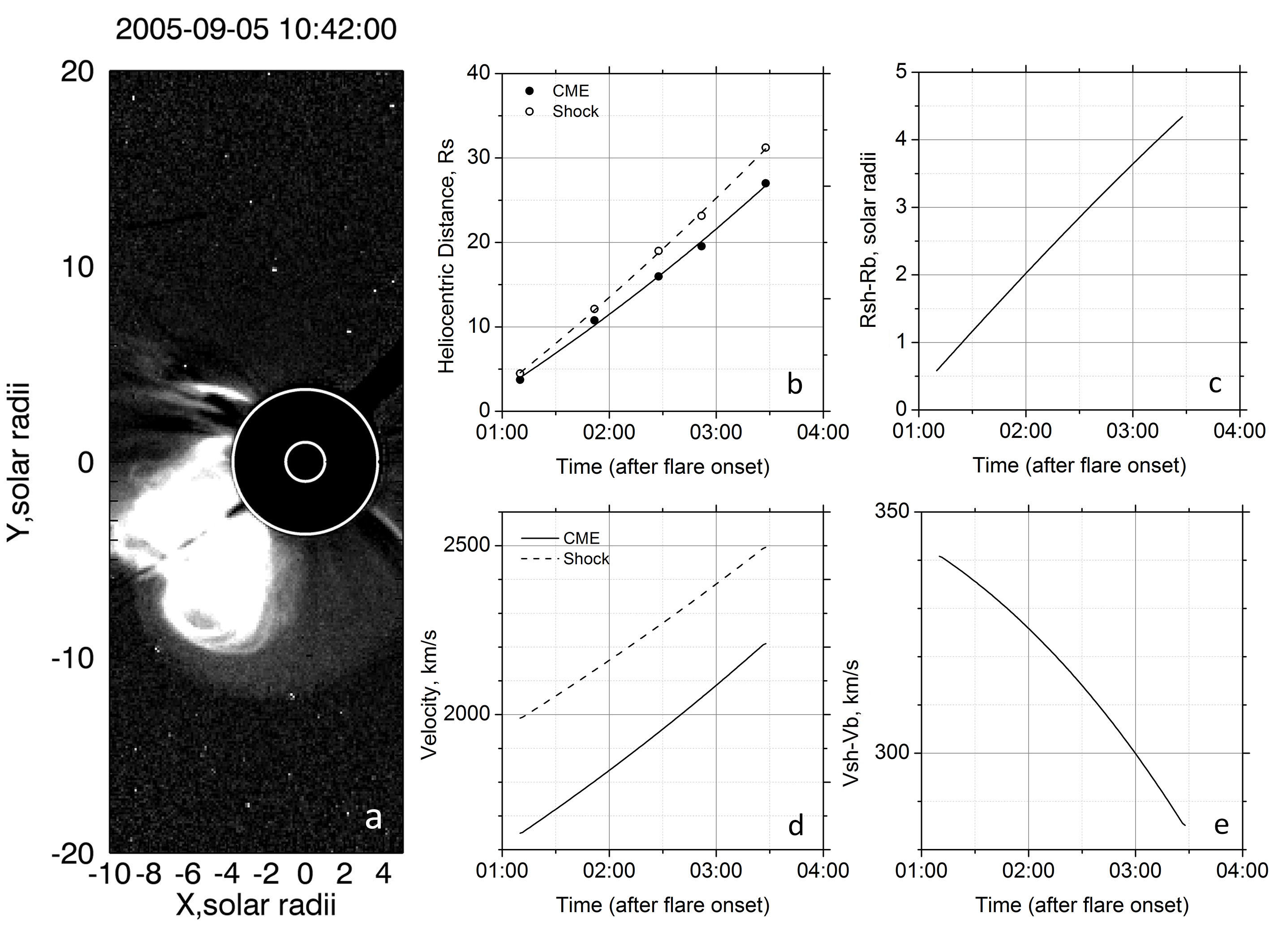}}
    \caption{(a) - LASCO C2 running difference image; (b) - The shock and CME body position was determined along the PA=150$^{\circ}$ direction; (c) - difference in positions of the shock and CME body depending on time (the shown line is described with the equation that was derived from difference of equations describing the shock and CME body regression lines); (d) – time dependence of the shock (upper line) and CME body velocity (lower line); (e) – velocity difference of the shock and CME body depending on time. In the plots, time track starts from the flare onset }
\end{figure}

Figure 5 shows all the same as Fig. 4, but for all the considered limb events. We can see that all behavior features of a shock and CME body position and velocity, which were revealed for the 27.07.2005 event, qualitatively are held, on average, for all the limb mass ejections considered. Unlike Fig. 4, this figure adds distance-dependences of difference in CME body and shock positions, and velocity differences of these structures.

\begin{figure}[!ht]
    {\includegraphics[width=1.\textwidth]{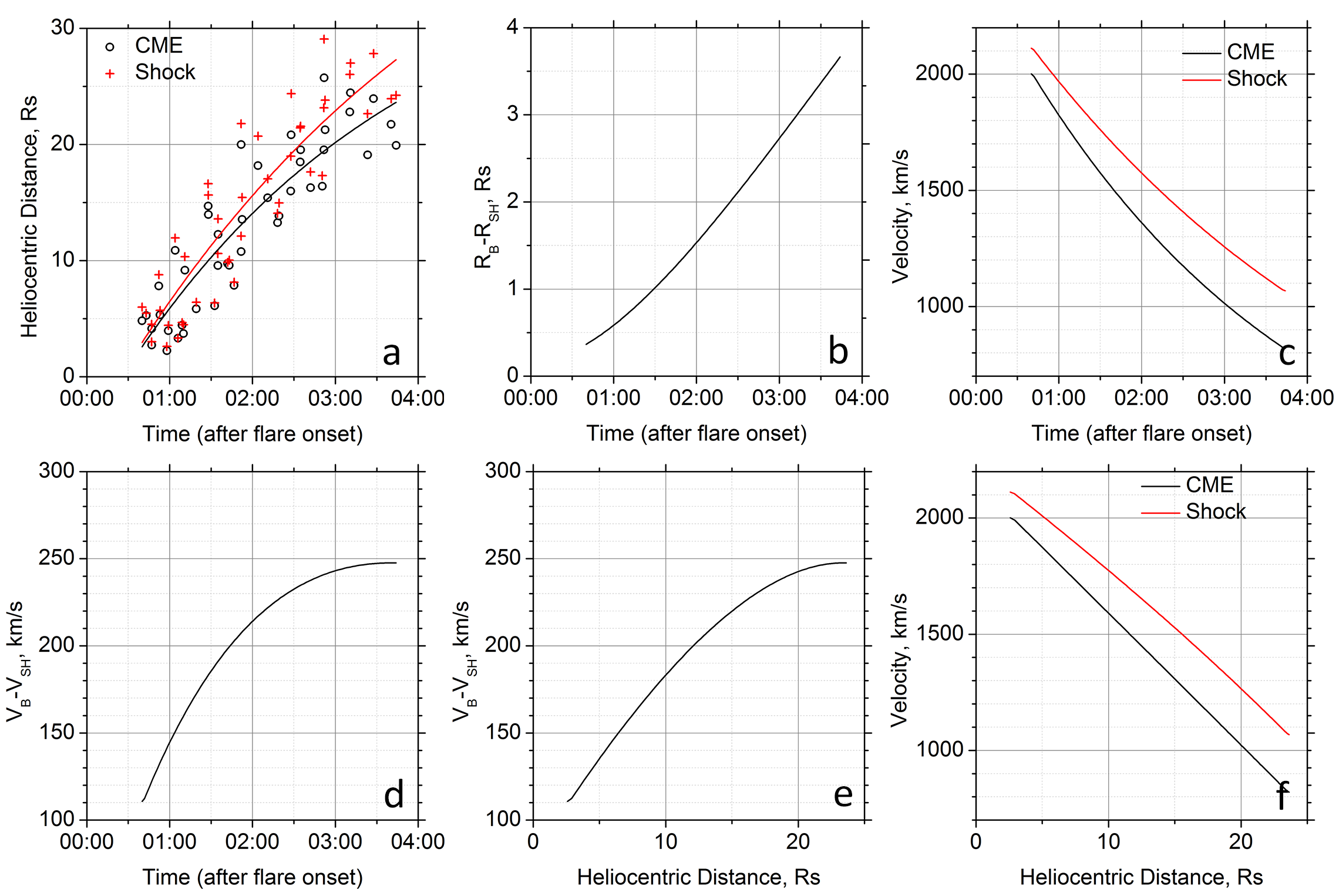}}
    \caption{Dependences describing shock and CME body kinematics for all the considered limb events, similar to dependences shown in Fig. 4. In (e,f) we added distance dependences of the difference in positions of shock and CME body boundary, and the difference in the velocities of shock and CME body.}
\end{figure}

Figure 6 is similar to Fig. 4, but it shows the results for the halo CME observed on 07.03.2012. Qualitatively, all the conclusions made in comparison of kinematics for the shock and limb CME registered on 27.07.2005 are the same for the halo CME. The only noticeable difference is that the velocity difference between CME body and shock varies slightly with time (distance) as compared to limb CME, although, as for limb CMEs, it decreases as CME moves on. In other words, expansion velocity of the CME body and associated shock changes with time in roughly the same manner.

\begin{figure}[!ht]
    {\includegraphics[width=1.\textwidth]{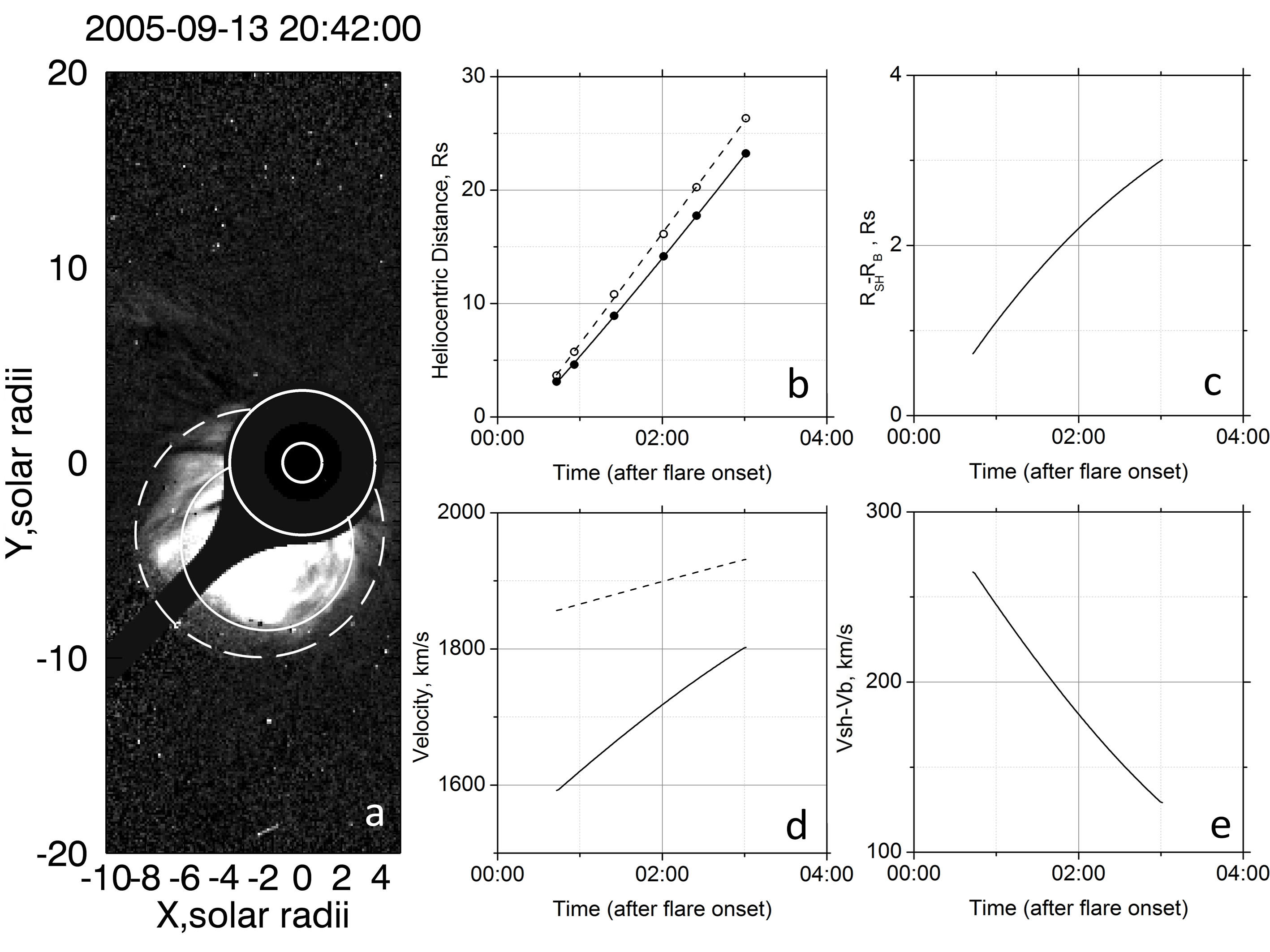}}
    \caption{Same as in Fig. 3, but for the halo CME registered on 2005-09-13.}
\end{figure}
Fig. 6. 

Figure 7 shows the results for all halo CMEs considered. On average, all the features of the body of halo-type CME and shock kinematics are similar to those of shock and body of the halo CME registered on 2005-09-13. At the same time, difference in velocities of the two structures is noticeably lower here than for the 2005-09-13 CME. Besides, unlike the 07.03.2012 event, for all halo CMEs considered, the shock velocity on average decreases with time much faster than the CME body velocity.  
Comparing plots in Fig. 7 with plots in Fig. 5, note that for all the timepoints considered from the beginning of CME body and shock registration in coronagraphs' FOV, the $R_{sh}/R_{\odot}$ - $R_b/R_{\odot}$ difference is larger for halo CMEs than for limb CMEs. In other words, distance from the shock flanks to the boundary of CME body is greater on average than toward the CME translatory movement. This is also seen for limb CMEs in the corona images if viewed transverse to the CME axis. Difference in velocities of CME body and shock is on average greater for halo CMEs compared to limb mass ejections. This means that the shock flanks are moving from CME body faster than the shock moves from CME body along the movement direction.   

\begin{figure}[!ht]
    {\includegraphics[width=1.\textwidth]{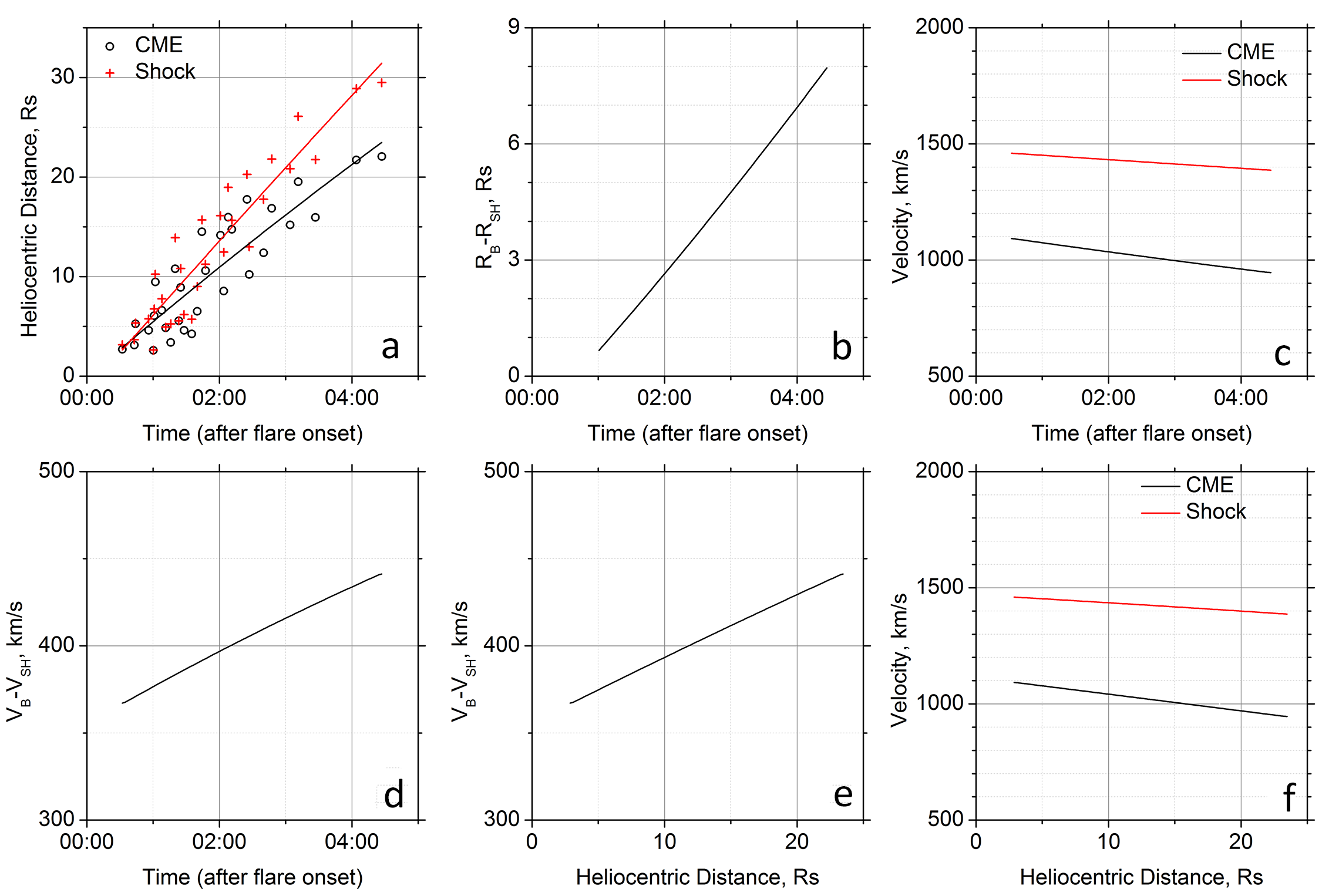}}
    \caption{Same as in Fig. 5, but for all halo CMEs considered.}
\end{figure}

Figure 8 demonstrates the results of comparing shock and CME body kinematics in 3D space for 5 events. We can see that in terms of quality, the results obtained in this case are consistent with the results for limb CMEs and halo CMEs. Still, quantitatively, the results obtained in this case differ from the results for limb CMEs and halo CMEs in the coronagraphs' plane of sky (Fig. 5, 7). First, the external boundary of a CME body and associated shock are registered at greater maximum distances, and they have greater maximum velocity that CMEs registered in the plane of the sky. Greater maximum differences in positions of CME body boundary and shock and maximum velocity differences are typical for them compared to CMEs observed in the sky plane. These differences are easily explained by the fact that the CME body and associated shock, which are observed in the coronagraph's FOV, are the projection of real CMEs and shocks in 3D space to the sky plane.

\begin{figure}[!ht]
    {\includegraphics[width=1.\textwidth]{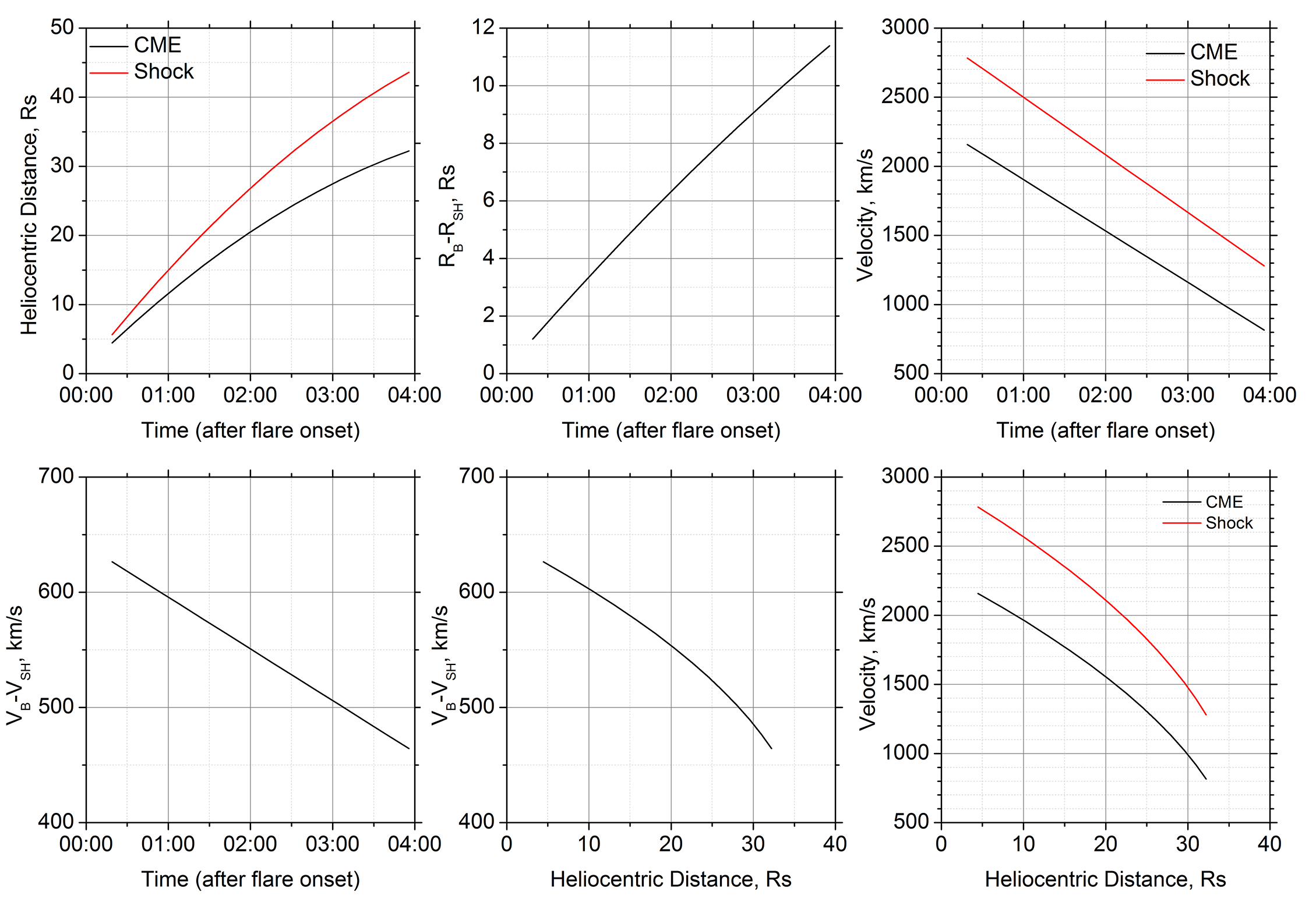}}
    \caption{Comparison of CME body and shock 3D kinematics obtained using the CME ``Ice cream cone model''. The meaning of plots in all panels is the same as in Figure 5.}
\end{figure}

\section{Conclusion}

Even visual analysis of CME and its associated shock movement in the coronagraphs' FOV leads us to the conclusion that kinematics of these two solar structures is different. Despite the fact that for two CMEs, CME and shock kinematic characteristics in FOV of STEREO/COR2 coronagraph were compared \citep{2011ApJ...736L...5M,2016AIPC.1714c0003V}, these results can hardly be deemed as comprehensive ones for the following reasons: first, not enough events were studied, which does not allow making statistically significant conclusions; second, we need to specifically explore and then compare the results for two series of CMEs: limb CMEs and halo CMEs with their sources near the center of the solar disk. This is related to the fact that limb CMEs take part in two types of movement simultaneously: they move translatory as a whole and expand, including expansion toward the direction of the movement. While for halo CMEs with the sources near the center of solar disk in coronagraph's FOV, we can mainly observe their transverse extension, only the shock flanks are registered in the process. It will be also helpful to compare all the obtained results of coronagraph FOV observations with the results available for CME and shock movement in 3D space. 
In this paper, based on LASCO C2 and C3 data, depending on time (distance), we have determined positions and velocities of fast (linear projection velocity over 1500 km/s) limb CME body with their sources near the limb, and the body of halo-type CME with the sources near the solar disk center. These CME body parameters have been compared to similar kinematic characteristics of their associated shock waves. For halo CMEs with their sources near the solar disk center and their associated shocks, we determined and compared their kinematic characteristics in 3D space. The feature of our analysis of white corona images is that for most of the analyzed events we were able to observe shock as a jump of brightness; the shock front was observed as a feature on the radial distribution of corona brightness. 
It has been shown that both for limb CMEs and halo CMEs, the projection velocity of the CME body in the plane of sky at each timepoint is less than the shock velocity and, on average, for all the events considered, both velocities go down with time (distance) as the mass ejection moves, and difference in positions of a shock wave and CME body increases with time (distance). In the process, the shock velocity decreased with time faster than the CME body velocity, with time, this results in decreased difference of the two structures' velocities. It was revealed that for halo CME, the rate of velocity change with time (distance) for both CME body and shock is higher than for limb CME.
We compared kinematic characteristics of the body of halo-type CME and associated shock in coronagraph's FOV with the characteristics of CME and shock movement in 3D space using the “Ice-cream cone model” for CME and shock. It was established that in 3D space, behavior of the body of halo-type CME and shock kinematic characteristics is described with the same features as for limb CMEs and halo CMEs in FOV of LASCO C2 and С3 coronagraphs. At the same time, normally, there is a slight change in the difference between the CME body and associated shock with time compared to the observation in the plane of the sky. Analysis of CME body and associated shock motion in 3D space has shown that in this case we can observe these structures moving away from the Sun for great distances, and they move with high velocities. Difference  of the CME body boundary and shock wave positions, and of their velocities is also greater for the model CME in 3D space than these parameters of CME in the plane of the sky. The listed differences in CME body and shock kinematic parameters are related to the fact that in the plane of the sky we measure projection parameters of two structures (position and velocity), which are less than for structures in three dimensions.

The authors thank the LASCO team and the authors of \url{https://cdaw.gsfc.nasa.gov/CME_list/} and \url{https://cdaw.gsfc.nasa.gov/CME_list/HALO/halo.html} for the opportunity of free use of LASCO data and their catalogues. The paper is prepared with the support of RFBR grants No. 15-02-01077-a and No. 16-32-00315, it is also a part of the Research and Development plan of ISTP SB RAS for 2016-2018. II.16.1.6. "Geo-effectie processes in chromosphere and solar corona" (Basic project).

\bibliographystyle{model3-num-names}
\bibliography{biblio}

\begin{thebibliography}{14}
\providecommand{\natexlab}[1]{#1}
\providecommand{\url}[1]{\texttt{#1}}
\providecommand{\href}[2]{#2}
\providecommand{\path}[1]{#1}
\providecommand{\eprint}[1]{\href{http://arxiv.org/abs/#1}{\path{#1}}}
\providecommand{\DOIprefix}{doi:}
\providecommand{\ArXivprefix}{arXiv:}
\providecommand{\URLprefix}{URL: }
\providecommand{\Pubmedprefix}{pmid:}
\providecommand{\doi}[1]{\href{http://dx.doi.org/#1}{\path{#1}}}
\providecommand{\Pubmed}[1]{\href{pmid:#1}{\path{#1}}}
\providecommand{\BIBand}{and}
\providecommand{\bibinfo}[2]{#2}
\ifx\xfnm\undefined \def\xfnm[#1]{\unskip,\space#1}\fi
\bibitem[{{Brueckner} et~al.(1995){Brueckner}, {Howard}, {Koomen}, {Korendyke},
  {Michels}, {Moses} et~al.}]{1995SoPh..162..357B}
\bibinfo{author}{{Brueckner}\xfnm[ G.E.]}, \bibinfo{author}{{Howard}\xfnm[
  R.A.]}, \bibinfo{author}{{Koomen}\xfnm[ M.J.]},
  \bibinfo{author}{{Korendyke}\xfnm[ C.M.]}, \bibinfo{author}{{Michels}\xfnm[
  D.J.]}, \bibinfo{author}{{Moses}\xfnm[ J.D.]}, et~al.
\newblock \bibinfo{title}{{The Large Angle Spectroscopic Coronagraph (LASCO)}}.
\newblock \bibinfo{journal}{\solphys}
  \bibinfo{year}{1995};\bibinfo{volume}{162}:\bibinfo{pages}{357--402}.
\newblock \DOIprefix\doi{10.1007/BF00733434}.
\bibitem[{{Domingo} et~al.(1995){Domingo}, {Fleck} and
  {Poland}}]{1995SoPh..162....1D}
\bibinfo{author}{{Domingo}\xfnm[ V.]}, \bibinfo{author}{{Fleck}\xfnm[ B.]},
  \bibinfo{author}{{Poland}\xfnm[ A.I.]}.
\newblock \bibinfo{title}{{The SOHO Mission: an Overview}}.
\newblock \bibinfo{journal}{\solphys}
  \bibinfo{year}{1995};\bibinfo{volume}{162}:\bibinfo{pages}{1--37}.
\newblock \DOIprefix\doi{10.1007/BF00733425}.
\bibitem[{{Fainshtein} and {Egorov}(2015)}]{Fainshtein2015}
\bibinfo{author}{{Fainshtein}\xfnm[ V.G.]}, \bibinfo{author}{{Egorov}\xfnm[
  Y.I.]}.
\newblock \bibinfo{title}{{Initiation of CMEs associated with filament
  eruption, and the nature of CME related shocks}}.
\newblock \bibinfo{journal}{Advances in Space Research}
  \bibinfo{year}{2015};\bibinfo{volume}{55}:\bibinfo{pages}{798--807}.
\newblock \DOIprefix\doi{10.1016/j.asr.2014.05.019}.
\bibitem[{{Gopalswamy} and {Yashiro}(2011)}]{Gopalswamy2011}
\bibinfo{author}{{Gopalswamy}\xfnm[ N.]}, \bibinfo{author}{{Yashiro}\xfnm[
  S.]}.
\newblock \bibinfo{title}{{The Strength and Radial Profile of the Coronal
  Magnetic Field from the Standoff Distance of a Coronal Mass Ejection-driven
  Shock}}.
\newblock \bibinfo{journal}{\apjl}
  \bibinfo{year}{2011};\bibinfo{volume}{736}:\bibinfo{eid}{L17}.
\newblock \DOIprefix\doi{10.1088/2041-8205/736/1/L17}.
  \href{http://arxiv.org/abs/1106.4832}{\tt arXiv:1106.4832}.
\bibitem[{{Howard} et~al.(2008){Howard}, {Moses}, {Vourlidas}, {Newmark},
  {Socker}, {Plunkett} et~al.}]{stereo}
\bibinfo{author}{{Howard}\xfnm[ R.A.]}, \bibinfo{author}{{Moses}\xfnm[ J.D.]},
  \bibinfo{author}{{Vourlidas}\xfnm[ A.]}, \bibinfo{author}{{Newmark}\xfnm[
  J.S.]}, \bibinfo{author}{{Socker}\xfnm[ D.G.]},
  \bibinfo{author}{{Plunkett}\xfnm[ S.P.]}, et~al.
\newblock \bibinfo{title}{Sun earth connection coronal and heliospheric
  investigation (secchi)}.
\newblock \bibinfo{journal}{\ssr}
  \bibinfo{year}{2008};\bibinfo{volume}{136}:\bibinfo{pages}{67--115}.
\newblock \DOIprefix\doi{10.1007/s11214-008-9341-4}.
\bibitem[{{Kaiser} et~al.(2008){Kaiser}, {Kucera}, {Davila}, {St.~Cyr},
  {Guhathakurta} and {Christian}}]{2008SSRv..136....5K}
\bibinfo{author}{{Kaiser}\xfnm[ M.L.]}, \bibinfo{author}{{Kucera}\xfnm[ T.A.]},
  \bibinfo{author}{{Davila}\xfnm[ J.M.]}, \bibinfo{author}{{St.~Cyr}\xfnm[
  O.C.]}, \bibinfo{author}{{Guhathakurta}\xfnm[ M.]},
  \bibinfo{author}{{Christian}\xfnm[ E.]}.
\newblock \bibinfo{title}{{The STEREO Mission: An Introduction}}.
\newblock \bibinfo{journal}{\ssr}
  \bibinfo{year}{2008};\bibinfo{volume}{136}:\bibinfo{pages}{5--16}.
\newblock \DOIprefix\doi{10.1007/s11214-007-9277-0}.
\bibitem[{{Maloney} and {Gallagher}(2011)}]{2011ApJ...736L...5M}
\bibinfo{author}{{Maloney}\xfnm[ S.A.]}, \bibinfo{author}{{Gallagher}\xfnm[
  P.T.]}.
\newblock \bibinfo{title}{{STEREO Direct Imaging of a Coronal Mass
  Ejection-driven Shock to 0.5 AU}}.
\newblock \bibinfo{journal}{\apjl}
  \bibinfo{year}{2011};\bibinfo{volume}{736}:\bibinfo{eid}{L5}.
\newblock \DOIprefix\doi{10.1088/2041-8205/736/1/L5}.
  \href{http://arxiv.org/abs/1106.1593}{\tt arXiv:1106.1593}.
\bibitem[{{Ontiveros} and {Vourlidas}(2009)}]{Ontiveros2009}
\bibinfo{author}{{Ontiveros}\xfnm[ V.]}, \bibinfo{author}{{Vourlidas}\xfnm[
  A.]}.
\newblock \bibinfo{title}{{Quantitative Measurements of Coronal Mass
  Ejection-Driven Shocks from LASCO Observations}}.
\newblock \bibinfo{journal}{\apj}
  \bibinfo{year}{2009};\bibinfo{volume}{693}:\bibinfo{pages}{267--275}.
\newblock \DOIprefix\doi{10.1088/0004-637X/693/1/267}.
  \href{http://arxiv.org/abs/0811.3743}{\tt arXiv:0811.3743}.
\bibitem[{{Savani} et~al.(2012){Savani}, {Shiota}, {Kusano}, {Vourlidas} and
  {Lugaz}}]{2012ApJ...759..103S}
\bibinfo{author}{{Savani}\xfnm[ N.P.]}, \bibinfo{author}{{Shiota}\xfnm[ D.]},
  \bibinfo{author}{{Kusano}\xfnm[ K.]}, \bibinfo{author}{{Vourlidas}\xfnm[
  A.]}, \bibinfo{author}{{Lugaz}\xfnm[ N.]}.
\newblock \bibinfo{title}{{A Study of the Heliocentric Dependence of Shock
  Standoff Distance and Geometry using 2.5D Magnetohydrodynamic Simulations of
  Coronal Mass Ejection Driven Shocks}}.
\newblock \bibinfo{journal}{\apj}
  \bibinfo{year}{2012};\bibinfo{volume}{759}:\bibinfo{eid}{103}.
\newblock \DOIprefix\doi{10.1088/0004-637X/759/2/103}.
  \href{http://arxiv.org/abs/1209.1990}{\tt arXiv:1209.1990}.
\bibitem[{{Volpes} and {Bothmer}(2016)}]{2016AIPC.1714c0003V}
\bibinfo{author}{{Volpes}\xfnm[ L.]}, \bibinfo{author}{{Bothmer}\xfnm[ V.]}.
\newblock \bibinfo{title}{{Determining CME-driven shock parameters from remote
  sensing observations}}.
\newblock In: \bibinfo{booktitle}{American Institute of Physics Conference
  Series}; vol. \bibinfo{volume}{1714} of \emph{\bibinfo{series}{American
  Institute of Physics Conference Series}}. \bibinfo{year}{2016}, p.
  \bibinfo{pages}{030003}.
\newblock \DOIprefix\doi{10.1063/1.4942572}.
\bibitem[{{Vourlidas} and {Bemporad}(2012)}]{Vourlidas2012}
\bibinfo{author}{{Vourlidas}\xfnm[ A.]}, \bibinfo{author}{{Bemporad}\xfnm[
  A.]}.
\newblock \bibinfo{title}{{A decade of coronagraphic and spectroscopic studies
  of CME-driven shocks}}.
\newblock In: \bibinfo{editor}{{Heerikhuisen}\xfnm[ J.]},
  \bibinfo{editor}{{Li}\xfnm[ G.]}, \bibinfo{editor}{{Pogorelov}\xfnm[ N.]},
  \bibinfo{editor}{{Zank}\xfnm[ G.]}, editors. \bibinfo{booktitle}{American
  Institute of Physics Conference Series}; vol. \bibinfo{volume}{1436} of
  \emph{\bibinfo{series}{American Institute of Physics Conference Series}}.
  \bibinfo{year}{2012}, p. \bibinfo{pages}{279--284}.
\newblock \DOIprefix\doi{10.1063/1.4723620}.
  \href{http://arxiv.org/abs/1207.1603}{\tt arXiv:1207.1603}.
\bibitem[{{Vourlidas} and {Ontiveros}(2009)}]{Vourlidas2009}
\bibinfo{author}{{Vourlidas}\xfnm[ A.]}, \bibinfo{author}{{Ontiveros}\xfnm[
  V.]}.
\newblock \bibinfo{title}{{A Review of Coronagraphic Observations of Shocks
  Driven by Coronal Mass Ejections}}.
\newblock In: \bibinfo{editor}{{Ao}\xfnm[ X.]},
  \bibinfo{editor}{{Burrows}\xfnm[ G.Z.R.]}, editors.
  \bibinfo{booktitle}{American Institute of Physics Conference Series}; vol.
  \bibinfo{volume}{1183} of \emph{\bibinfo{series}{American Institute of
  Physics Conference Series}}. \bibinfo{year}{2009}, p.
  \bibinfo{pages}{139--146}.
\newblock \DOIprefix\doi{10.1063/1.3266770}.
  \href{http://arxiv.org/abs/0908.1996}{\tt arXiv:0908.1996}.
\bibitem[{{Vourlidas} et~al.(2003){Vourlidas}, {Wu}, {Wang}, {Subramanian} and
  {Howard}}]{Vourlidas2003}
\bibinfo{author}{{Vourlidas}\xfnm[ A.]}, \bibinfo{author}{{Wu}\xfnm[ S.T.]},
  \bibinfo{author}{{Wang}\xfnm[ A.H.]}, \bibinfo{author}{{Subramanian}\xfnm[
  P.]}, \bibinfo{author}{{Howard}\xfnm[ R.A.]}.
\newblock \bibinfo{title}{{Direct Detection of a Coronal Mass
  Ejection-Associated Shock in Large Angle and Spectrometric Coronagraph
  Experiment White-Light Images}}.
\newblock \bibinfo{journal}{\apj}
  \bibinfo{year}{2003};\bibinfo{volume}{598}:\bibinfo{pages}{1392--1402}.
\newblock \DOIprefix\doi{10.1086/379098}.
  \href{http://arxiv.org/abs/astro-ph/0308367}{\tt arXiv:astro-ph/0308367}.
\bibitem[{{Xue} et~al.(2005){Xue}, {Wang} and {Dou}}]{Xue2005}
\bibinfo{author}{{Xue}\xfnm[ X.H.]}, \bibinfo{author}{{Wang}\xfnm[ C.B.]},
  \bibinfo{author}{{Dou}\xfnm[ X.K.]}.
\newblock \bibinfo{title}{{An ice-cream cone model for coronal mass
  ejections}}.
\newblock \bibinfo{journal}{Journal of Geophysical Research (Space Physics)}
  \bibinfo{year}{2005};\bibinfo{volume}{110}:\bibinfo{eid}{A08103}.
\newblock \DOIprefix\doi{10.1029/2004JA010698}.

\end{thebibliography}


\end{document}